\documentclass[sigconf,natbib=false]{acmart}

\AtBeginDocument{%
 }

\copyrightyear{2024}
\acmYear{2024}
\setcopyright{rightsretained}
\acmConference[SIGMOD-Companion '24]{Companion of the 2024 International Conference on Management of Data}{June 9--15, 2024}{Santiago, AA, Chile}
\acmBooktitle{Companion of the 2024 International Conference on Management of Data (SIGMOD-Companion '24), June 9--15, 2024, Santiago, AA, Chile}
\acmDOI{10.1145/3626246.3653386}
\acmISBN{979-8-4007-0422-2/24/06}

\makeatletter
\gdef\@copyrightpermission{
  \begin{minipage}{0.3\columnwidth}
   \href{https://creativecommons.org/licenses/by/4.0/}{\includegraphics[width=0.90\textwidth]{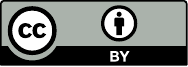}}
  \end{minipage}\hfill
  \begin{minipage}{0.7\columnwidth}
   \href{https://creativecommons.org/licenses/by/4.0/}{This work is licensed under a Creative Commons Attribution International 4.0 License.}
  \end{minipage}
  \vspace{5pt}
}
\makeatother

\usepackage{balance}
\usepackage{alltt}
\usepackage{multirow}
\usepackage{fancyvrb}

\RequirePackage[
  datamodel=acmdatamodel,
  style=acmnumeric,
  ]{biblatex}

\begin{document}

\title{PG-Triggers: Triggers for Property Graphs}

\author{Stefano Ceri}
\orcid{0000-0003-0671-2415}
\affiliation{%
  \institution{Politecnico di Milano}
  \city{Milan}
  \country{Italy}
}
\email{stefano.ceri@polimi.it}

\author{Anna Bernasconi}
\orcid{0000-0001-8016-5750}
\affiliation{%
  \institution{Politecnico di Milano}
  \city{Milan}
  \country{Italy}
}
\email{anna.bernasconi@polimi.it}

\author{Alessia Gagliardi}
\affiliation{%
  \institution{Politecnico di Milano}
  \city{Milan}
  \country{Italy}
}
\email{alessia1.gagliardi@mail.polimi.it}

\author{Davide Martinenghi}
\affiliation{%
  \institution{Politecnico di Milano}
  \city{Milan}
  \country{Italy}
}
\email{davide.martinenghi@polimi.it}

\author{Luigi Bellomarini}
\affiliation{%
  \institution{Banca d'Italia}
  \city{Rome}
  \country{Italy}
}
\email{luigi.bellomarini@bancaditalia.it}

\author{Davide Magnanimi}
\affiliation{%
  \institution{Banca d'Italia \& Politecnico di Milano}
  \city{Rome \& Milan}
  \country{Italy}
}
\email{davide.magnanimi@bancaditalia.it}

\renewcommand{\shortauthors}{Stefano Ceri et al.}

\begin{abstract}
Graph databases are emerging as the leading data management technology for storing large knowledge graphs; significant efforts are ongoing to produce new standards (such as the Graph Query Language, GQL), as well as enrich them with properties, types, schemas, and keys. In this article, we introduce PG-Triggers, a complete proposal for adding triggers to Property Graphs, along the direction marked by the SQL3 Standard. 

We define the syntax and semantics of PG-Triggers and then illustrate how they can be implemented on top of Neo4j, one of the most popular graph databases. In particular, we introduce a syntax-directed translation from PG-Triggers into Neo4j, which makes use of the so-called {\it APOC triggers}; APOC is a community-contributed library for augmenting the Cypher query language supported by Neo4j.  We also cover Memgraph, and show that our approach applies to this system in a similar way. We illustrate the use of PG-Triggers through a life science application inspired by the COVID-19 pandemic. 

The main objective of this article is to introduce an active database standard for graph databases as a first-class citizen at a time when reactive graph management is in its infancy, so as to minimize the conversion efforts towards a full-fledged standard proposal.
\end{abstract}




\maketitle

\section{Introduction}

Graph databases are becoming increasingly important as frameworks for representing and understanding the intricate connections that exist in the real world~\cite{sakr2021future}. 
Thanks to their expressive query languages, 
rich customer support, and strong performance, they are steadily more used to store large knowledge graphs, in a variety of domains that include, e.g., mobility, social, and biological networks.

As customary in data management evolution, standards for graph databases are
emerging, most importantly the Graph Query Language (GQL)~\cite{greengqlScope}, whose roadmap is followed with interest by the major companies in the field. In addition,
the research community has proposed various 
formalizations so as to enrich the semantics of graph databases, first by shaping them in the form of \emph{Property Graphs}~\cite{Bonifati2018}, and then by defining the notions of \emph{PG-Keys}~\cite{angles2021pg} and \emph{PG-Schema}~\cite{bonifati2022pg}. 

In this paper, we follow up along this trend and propose \emph{PG-Triggers}. Triggers exist
since the birth of relational databases~\cite{eswaran}, have been studied in~\cite{widom1995active}, and formalized in the ISO-ANSI SQL3 Standard~\cite{melton2001sql}. So far, they have not been formalized by the graph database research community, although they can be informally supported by most graph database systems, even if in diversified ways (see Section~\ref{sec:relwork}). Hence, our proposal for PG-Triggers has the potential to influence future standard development as well as suggest new directions to the evolution of graph databases.

We demonstrate our idea in practice by focusing on Neo4j -- one of the most representative graph database systems. We show that Neo4j already supports all the required components for implementing our PG-Trigger concepts; however, it does so within 
a community-supported library, called \emph{Awesome Procedures on Cypher} (APOC)~\cite{apoc}, and not as part of Cypher~\cite{francis2018cypher} -- the declarative graph query language adopted by Neo4j. We show a syntax-directed translation of PG-Triggers into APOC triggers, while also discussing some intricacies that must be solved for an effective translation. Moreover, we illustrate several limitations of APOC Neo4j triggers. We also cover Memgraph, the only other graph database that offers trigger support, and show that our approach easily extends to Memgraph, although with small differences. The purpose of translations is not to advocate their use in response to an established standard along our PG-Trigger proposal - as we would instead expect each system to directly support the standard. Our translation schemes show that current Neo4j or Memgraph solutions could rather easily evolve into unified, orthogonal, clear, and solid abstractions, along the PG-Trigger proposal, so as to favor a wider use of empowered triggers. Indeed, based on past experience with commercial relational systems \cite{widom1995active-commercial}, this is the right time to look for convergence among graph databases.

Our proposal is then applied to build a reactive model for a life science application addressing critical aspects of the COVID-19 pandemic. Starting from our CoV2K knowledge base~\cite{alfonsi2022cov2k}, we show a fragment of the knowledge base modeled using property graphs and implemented using Neo4j, and then several PG-Triggers that define a reactive behavior, in particular by responding to events such as the discovery of critical SARS-CoV-2 mutations, or the diffusion of unknown variants, or the increase of hospitalizations requiring intensive care treatment. In this way, 
after providing the foundations of PG-Triggers, our work establishes the first brick in the development of reactive processing over graphs, covering complex scenarios like those occurring in the pandemic setting.

\smallskip

\noindent \textbf{Contributions.}
Our main contribution is a proposal for adding reactive behavior in the form of triggers to graph databases. This proposal aims to be natural and useful:
\begin{itemize}
\item Natural, as it suitably adapts the recommendations of the SQL3 standard to a graph setting, thereby adhering to the principle of least surprise, for the great benefit of people already acquainted with the well-known corresponding relational notions.
\item Useful, as knowledge management through reactive behavior proves to be very effective in numerous \textit{knowledge-intensive} (instead of just data-intensive) scenarios.
\end{itemize}
The adaptation of the notion of triggers to Property Graphs is, however, far from trivial, as, on one hand, it requires dealing with non-relational concepts such as nodes, relationships, labels, and properties, and, on the other hand, it lacks a full correspondence with the notion of table, which, in the relational case, is the source of events that can be monitored by triggers. To this end, our proposal identifies the notion of label as the most suitable and natural choice for defining a set of target elements in the graph, much in the same way in which tables do in SQL triggers. 

In order to demonstrate our solution, we show how PG-Triggers can be implemented on top of Neo4j through the APOC trigger library. While this implementation provides evidence of the feasibility of our proposal in the currently most popular graph database, it also highlights the inadequacy of Neo4j APOC triggers: not only are they unstable (we experienced several changes during the development of our implementation), but they also lack a few important ingredients that proved extremely useful in our examples, among which the support for a correct cascading of triggers (occurring when a trigger's action causes the activation of other triggers), for event-specific triggering action times, and for instance-level vs set-level trigger granularities.
We support these features in our PG-Triggers proposal, whose syntax and semantics are streamlined as much as possible, so as to combine ease of use with expressivity, in the hopes that our proposal can drive forthcoming standardization choices in Property Graphs.

\smallskip

\noindent\textbf{Outline.} After discussing previous standardization attempts and related work in Section~\ref{sec:relwork}, we offer a comparative review of graph database technology in Section~\ref{sec:comp-rev}, by also surveying the different levels of support for trigger-like constructs in current systems. Our proposed syntax and semantics for PG-Triggers are illustrated in Section~\ref{sec:pgtriggers}, while their implementation using the APOC library is discussed in Section~\ref{sec:mapping}; we also show, by difference, a very similar implementation in Memgraph. 
We exemplify our proposal in Section~\ref{sec:example} by providing reactive support to a life science application tackling COVID-19. Finally, we discuss the extent and impact of PG-Triggers in Section~\ref{sec:conclusion}.

\section{Related Work}
\label{sec:relwork}

The Property Graph data model applies to a directed graph where nodes and edges are labeled, and each can have associated ⟨property, value⟩ pairs. The Property Graph data model has gained significant popularity and adoption in various graph database systems. Examples of systems that leverage this model include 
Neo4j~\cite{neo4j}, 
Memgraph~\cite{memgraph},
JanusGraph~\cite{janusGraph}, 
Amazon Neptune~\cite{neptune},
Nebula Graph~\cite{nebulaGraph},
TigerGraph~\cite{tigerGraph},
and more.
This large participation and attention on Property Graphs led to the idea of creating a standalone Property Graph query language to complement SQL, which was raised by ISO SC32/ WG3 members in early 2017 and is echoed in the GQL manifesto of May 2018~\cite{GQLManifesto}.

\textbf{The LDBC Graph Query Working Groups.} The Linked Data Benchmark Council (LDBC) Groups~\cite{WorkingGroups} are collaborative efforts aimed at advancing the state of the art in graph query processing, by bringing together researchers, industry experts, and practitioners. They promote standards and develop benchmarks for graph data management systems. The Graph Query Working Groups specifically focus on addressing challenges related to querying large-scale graph datasets efficiently. There are multiple subgroups, each focusing on a specific aspect of graph query processing. Among them, working groups for the LDBC Extended GQL Schema (LEX), the Property Graph Schema, the Existing Languages, and the Formal Semantics.

\textbf{G-CORE.} The authors of~\cite{angles2018g} present the syntax and semantics of the Graph Query Language Core (G-CORE), which supports graph pattern matching, property and structural filtering, aggregations, and path expressions. G-CORE incorporates graph-specific features while maintaining a close relationship with traditional relational algebra, making it accessible to both graph database practitioners and researchers. The advantages of G-CORE include its expressive power, formal semantics, and compatibility with existing graph database systems. 

\textbf{PG-Keys.} Along the direction of adding semantics to Property Graphs, the notion of key was then proposed for graph databases. PG-Keys~\cite{angles2021pg} are unique identifiers assigned to arbitrary subsets of nodes and edges within a Property Graph database. By assigning unique keys, different entities and relationships can be distinguished, preventing duplicates and maintaining data integrity. 

\textbf{PG-Schema.} The PG-Schema proposal~\cite{bonifati2022pg} introduces schemas for Property Graph databases; it addresses the need for a standardized approach to schema management, enabling users to define and enforce data constraints, specify relationships, and establish a clear structure for their graph data. 
PG-Schema uses the notion of PG-Type
for defining node and edge types, then
expresses type hierarchies and integrity constraints, also including PG-Keys.
The article presents a comprehensive framework for defining and evolving graph schemas, including support for schema inference, data validation, and schema evolution. The benefits of utilizing such a schema include improved data quality, stronger query optimization potential, and enhanced data governance.

\subsection{Brief history of reactive extensions for other data models}

Active extensions have been designed and implemented for a variety of data models, 
throughout the fifty-year-long development of database technology. Extensions prior to 1996 are described in~\cite{widom1995active}, with a review of commercial relational systems~\cite{widom1995active-commercial}, and chapters dedicated to Postgres, Ariel, Starburst, A-RDL, Chimera, and Ode. In particular, trigger events and actions were added to O++, the object-oriented query language in use in Ode, developed at AT\&T~\cite{gehani1992event}, while formal models for integrating database objects and triggers are discussed in~\cite{ceri1991deriving}. 

Similar active extensions have been proposed for XML; in particular, Active-XML~\cite{abiteboul2002active} augments XML with reactive computations modeled as services, in the context of a peer-to-peer distributed model of computations. An encompassing approach to the design of database applications using objects, deductive and active rules, is in~\cite{ceri1997designing}.

Only a few reactive extensions are discussed for research prototypes using graph databases. Among them, 
GraphFlow~\cite{kankanamge2017graphflow} and TurboFlux~\cite{kim2018turboflux}; both systems support continuous matching over graphs that change over time, using incremental sub-matching algorithms (along the lines described in~\cite{fan2013incremental}).
Turboflux provides a subgraph matching system by employing a concise representation of intermediate results and proposes to resolve the problems of existing methods with continuous subgraph matching for each
update operation.
GraphFlow is an interesting system developed at the University of Waterloo; it proposes several clever ideas for stream management and introduces Cypher++ as an active extension of the Cypher language. In particular, Cypher++ adds continuous queries to reactive processing, as subgraphs are continuously matched against query patterns, and reactions take place when matches occur.

\section{Comparative Review of Graph Database Technology}
\label{sec:comp-rev}

We analyze some of the commercial graph database systems, highlighting how they support reactive computations. In particular, we first analyze products focused on supporting graph (or RDF) data and then we consider systems mixing graph data with other kinds of data, including relational and document/key-value data.


\subsection{Graph Databases}
\label{sec:graph}


\subsubsection{Graph Databases with trigger support}

\begin{itemize}
    \item {\bf Neo4j.} Neo4j is an open-source NoSQL native graph database; 
    it is the most widely adopted graph database~\cite{PGranking}. 
    It has introduced Cypher, a declarative language for querying and manipulating graph data, the \textit{de facto} standard in graph databases. Neo4j does not support triggers natively, however, triggers are included in APOC (Awesome Procedures on Cypher), a popular extension library for Neo4j.
    \item {\bf Memgraph.}
    Memgraph is another open-source graph database compatible with Neo4j, built for real-time streaming. It offers an implementation of triggers that supports the execution of any openCypher \cite{openCypher} query.
\end{itemize}
Later in this paper, after introducing PG-Triggers, we will show how they can be translated into APOC triggers and Memgraph triggers.

\subsubsection{Graph Databases with event listeners}
 
\begin{itemize}

    \item {\bf JanusGraph.} JanusGraph is an open-source, distributed graph database system built on the graph computing framework Apache TinkerPop; it is efficient in handling very large graphs with billions of vertices and edges. 
    In JanusGraph, triggers can be produced through the "JanusGraph Bus", a collection of configurable logs to which JanusGraph writes changes to the graph. 
    
    
    \item {\bf Dgraph.} Dgraph is an open-source, horizontally scalable, distributed graph database. It 
    provides a flexible query language for querying and manipulating data, called DQL. Dgraph provides the Dgraph Lambda framework~\cite{dgraphLambda}, which can be used to react to events through Typescript or Javascript functions.
    
    \item {\bf Amazon Neptune.} Amazon Neptune is a fully managed graph database service provided by Amazon Web Services. Amazon Neptune supports both the Apache TinkerPop Gremlin graph traversal language and the openCypher query language for the Property Graph data model. For the Resource Description Framework (RDF) data model, Neptune supports the standard open language SPARQL query language~\cite{sparql}. 
    Neptune uses Amazon Simple Notification Service (Amazon SNS) to provide notifications when a Neptune event occurs.

    \item {\bf Stardog.} Stardog is a commercial graph DBMS with a connected Enterprise Knowledge Graph platform and virtualization capability. It supports GraphQL for the graph store and SPARQL for the RDF store. It uses Java event handlers to capture changes in the data.

\end{itemize}

\subsubsection{Other Graph Databases}

 \begin{itemize}
 
    \item {\bf Nebula Graph.} NebulaGraph is an open-source distributed, linear scalable database supporting efficient graph patterns. It supports Cypher and implements its own nGQL language; it    
    does not support reactive aspects.

    \item {\bf TigerGraph.} TigerGraph is a commercial parallel graph computing platform; it provides a graph query language called GSQL, which allows user-defined functions and procedures. 
    It does not include reactive aspects. 

    \item {\bf GraphDB.} GraphDB is a commercial graph database and RDF store, with efficient reasoning support. Queries are accepted in GraphQL and SPARQL; triggers are not supported.

\end{itemize}

\begin{table}[t]
\resizebox{\linewidth}{!}{%
\begin{tabular}{lccl}
\toprule
 & \textbf{Tr-G} & \textbf{Tr-R} & \multicolumn{1}{c}{\textbf{Ev-L}}\\
 \toprule
\textbf{Neo4j~\cite{neo4j}} & \checkmark &  & - \\ 
\textbf{Memgraph~\cite{memgraph}} & \checkmark &  & - \\ 

\textbf{JanusGraph~\cite{janusGraph}} & - &  & \checkmark (JSBus)\\ 
\textbf{Dgraph\cite{dgraph}} & - &  & \checkmark (Lambda)\\ 
\textbf{Amazon Neptune~\cite{neptune}} & - &  & \checkmark (SNS)\\ 
\textbf{Stardog~\cite{stardog}} & - &  & \checkmark (Java)\\ 

\textbf{Nebula Graph~\cite{nebulaGraph}} & - &  & -\\ 
\textbf{TigerGraph~\cite{tigerGraph}} & - &  & -\\ 
\textbf{GraphDB~\cite{graphdb}} & - &  & -\\ 

\midrule

\textbf{Oracle Graph Database~\cite{oracle}} & - & \checkmark  & -\\
\textbf{Virtuoso~\cite{virtuoso}} & - & \checkmark  & -\\ 
\textbf{AgensGraph~\cite{agensGraph}} & - & \checkmark  & -\\ 

\midrule

\textbf{Microsoft Azure Cosmos DB~\cite{azure}} & - &  & \checkmark(JS)\\ 
\textbf{OrientDB~\cite{orientDB}} & - &  & \checkmark(Hooks)\\ 
\textbf{ArangoDB~\cite{arangoDB}} & - &  & \checkmark\\ 
\bottomrule
\end{tabular}
}
\caption{Comparison of graph databases, focused on their management of reactive aspects. We consider Trigger Support in Graph Data (Tr-G) and in Relational Data (Tr-R), and availability of Event Listener (Ev-L), which can be exploited for building reactive behaviors with the support of procedures managed outside of the graph database.}
\label{tab:comparison}
\end{table}

\subsection{Mixed Graph-Relational Systems}

Mixed graph-relational systems are built by integrating two engines: a graph database and a relational database. The graph system supports a graph (Cypher-like) query language, whereas the relational system supports SQL. Queries can be built by assembling Cypher and SQL statements, where the former operates on graph data and the latter operates on tables. Relational engines support triggers, compliant with the SQL3 standard, but these do not operate on graph data. 

\begin{itemize}

    \item {\bf Oracle Graph Database and Graph Analytics.} The Oracle Graph Database supports the Property Graph data model, enabling graph analytics capabilities. Oracle Graph Database integrates with Oracle's broader ecosystem, including its SQL-based data management 
    and triggers.  

    \item {\bf Virtuoso.} Virtuoso is a multi-model DBMS developed by OpenLink (available both in open-source and commercial editions), providing SQL, XML, and RDF data management in a single multithreaded process. The trigger support is provided in the relational DBMS.
    
    \item {\bf AgensGraph.} Agens Graph is an open-source tool whose graph system supports the Property Graph as well as the relational data model; the latter is built upon PostgreSQL.
    Taking advantage of relational technology, AgensGraph supports 
    triggers.

\end{itemize}
\subsection{Mixed Graph-Document Databases}

Mixed graph-document systems are built by integrating graph data with document bases.

\begin{itemize}

  \item {\bf Microsoft Azure Cosmos DB.}
  Cosmos DB is a globally distributed, horizontally scalable, multi-model database service, queried through graph database APIs (Gremlin) and document database APIs (Mongo DB or Cassandra). 
  A JavaScript-integrated query API can be used to write triggers, which are classified into pre-triggers (executed before modifying a database item) and post-triggers (after modifying a database item). These must be specified for each database operation where their execution is expected.

    \item {\bf OrientDB.} OrientDB is an open-source, multi-model database management system that combines characteristics of document databases and of graph databases. It employs a SQL-like query language called OrientSQL and the native graph query language Gremlin. In this context, triggers (renamed as Hooks) enable the triggering of actions in response to document creation, modification, or deletion. 
   
    \item {\bf ArangoDB.} ArangoDB is a multi-model, open-source database system that combines the features of document, key-value, and graph databases into a single platform. It provides a native graph querying language called AQL (ArangoDB Query Language) for efficient graph traversals and graph-based analytics. ArangoDB supports different events that can be monitored through a listener (AbstractArangoEventListener). 
    
\end{itemize}

\subsection{Comparison and Discussion}
Table~\ref{tab:comparison} offers a synoptic view of how graph databases, with their main extensions, support reactive computations, either directly through triggers or indirectly through event listeners. Note, however, that trigger support in mixed relational systems operates just upon tables, i.e., the relational component. This comparison shows that, although forms of reactive processing exist in many commercial graph databases, they are not yet well developed. 

In summary, as of today, native triggers on graph data are available just in Neo4j and Memgraph; moreover, Neo4j triggers are still supported within community-defined APOC libraries and are not part of the standard language. Many other graph databases and hybrid systems support ingredients for building reactive systems (e.g., Hooks of OrientDB) but they are far away from supporting full-fledge trigger systems. This is the ideal time for setting the requirements for a trigger standard, so as to avoid misaligned deployment of similar but not identical features; we will show that signs of such misalignment exist already by focusing on Neo4j and Memgraph, the two graph databases with stronger trigger support, as well as market leaders within graph database according to \cite{PGranking}.

Note that a trigger standard is quite parsimonious, as triggers are compositions of event-condition-action rules that base all their syntax and semantics on a limited number of ingredients, which define: 
when they should be activated after a data creation, modification, or deletion; 
when their condition should be considered; and 
when their action should be executed if the condition is true. 
The essence of a trigger system should be as much as possible agnostic to the detailed aspects of the underlying query language - be it SQL, Cypher, or GQL. 
Fixing these aspects ahead of language standardization may lead to convergence -- at this time being inexpensive for graph database vendors, whereas -at the time of the SQL3 trigger standard- the development to obtain convergence among relational databases was quite hard \cite{widom1995active-commercial}.




\section{PG-Triggers Definition}
\label{sec:pgtriggers}

We next propose PG-Triggers, by discussing their syntax and semantics.

\begin{figure}[ht]
\resizebox{0.9\linewidth}{!}{
\begin{tabular}{l}
\texttt{CREATE TRIGGER <name> <time> <event>}\\
\texttt{ON <label>[.<property>]}\\
\texttt{[REFERENCING <alias for old or new>...]}\\
\texttt{FOR <granularity> <item>}\\
\texttt{[WHEN <condition>]}\\
\texttt{BEGIN}\\
\quad\texttt{<statement>}\\
\texttt{END}\\
\texttt{}\\
\texttt{<time> ::= \{ BEFORE | AFTER | ONCOMMIT | DETACHED \}}\\
\texttt{<event> ::= \{ CREATE | DELETE | SET | REMOVE \}}\\
\texttt{<granularity> ::= \{ EACH | ALL \}}\\
\texttt{<item> ::= \{ NODE | RELATIONSHIP \}}\\
\texttt{}\\
\texttt{<alias for old or new> ::=}\\
\quad\texttt{[OLDNODES | OLDRELS] AS <alias for old items> |}\\
\quad\texttt{[NEWNODES | NEWRELS] AS <alias for new items> |}\\
\quad\texttt{OLD AS <alias for old single item> |}\\
\quad\texttt{NEW AS <alias for new single item>} 
\end{tabular}
}
\caption{PG-Trigger Syntax} 
\label{fig:syntax}

\end{figure}

\subsection{Syntax}
\label{sec:syntax}

The PG-Triggers syntax, shown in Figure~\ref{fig:syntax}, is borrowed -- as much as possible -- from the SQL3 standard, as discussed, e.g., in Chapter 11 of~\cite{melton2001sql}. In our notation, 
upper case letters are reserved for terminal symbols;
nonterminals are in a low case and enclosed within $<>$ (angle brackets);
optionality is denoted by $[]$ (square brackets);
items of which only one is required are enclosed within braces $\{\}$;
alternatives are separated by the $|$ symbol;
and ellipsis points show that the preceding element can be repeated. 

Note that, due to the richness of the graph data model w.r.t. the relational model, the syntax has many more options.
In particular, note that graph items can either be nodes or relationships; we use a given label to select, out of all items, the specific set of items that is the trigger's \emph{target}.  
Note also that trigger events in graph databases are richer than those in relational databases, as they include the creation and deletion of nodes/relationships as well as the setting and removal of their labels and properties.
In analogy with the {\tt UPDATE} event of the SQL3 standard, the {\tt SET} and {\tt REMOVE} events can refer to properties; thus, the {\tt ON} clause may refer to labels (for nodes, relationships, and label themselves) but also to properties, identified by a {\tt <label>.<property>}) pair.

\subsection{Semantics}
\label{sub:trigger_semantics}
As usual, triggers include a \texttt{<condition>} predicate, 
which is considered at given action \texttt{<time>}(s), and a connected action \texttt{<statement>}, which is executed only if the corresponding condition predicate holds. We next describe the trigger semantics along the classical dimensions, by making explicit reference to their syntactic elements.

\begin{itemize}
\item {\bf Granularity.} We assume that each trigger execution is linked to a given query in a graph query language (GQL or Cypher); from our point of view, a query operates on an initial state and produces a final state, by creating or removing \texttt{<item>}s (i.e., nodes and relationships), or by setting or removing their labels and properties. These changes are considered at two possible levels of \texttt{<granularity>}: either individually ({\tt FOR EACH} clause) or collectively as a set ({\tt FOR ALL} clause).
\item {\bf Action Time.} As in relational databases, we consider triggers occurring {\tt BEFORE} and {\tt AFTER} the statement; in addition, we offer the {\tt ONCOMMIT} and {\tt DETACHED} option. As with relational triggers, {\tt BEFORE} statements should not produce arbitrary changes, but just condition {\tt NEW} states. {\tt ONCOMMIT} execution occurs before the commit execution, within the same transaction (possibly causing a rollback of the transaction as a whole), while {\tt DETACHED} execution takes place after a successful commit and operates within an autonomous transaction. Triggers that share the same action time must be ordered (see next).
\item {\bf Targeting.} Triggers in relational databases are targeted to tables. 
We opted for using labels as providers of an analogous context; therefore, in the {\tt ON} clause, we select as target the items with a particular \texttt{<label>}. 
\item {\bf Event Types.} Events refer to either nodes or relationships and include their creation or deletion, and the setting or removal of labels and properties. 
\item {\bf Transition Variables.} With individual-level granularity, {\tt OLD} and {\tt NEW} refer to the old and new state, respectively. With set-level granularity, transition variables are postfixed by the {\tt NODES} keyword or by the {\tt RELS} keyword; clearly, in this case, the item must be the same as in the \texttt{FOR} clause. Along with \cite{melton2001sql}, our proposed syntax offers an option for renaming transition variables through the \texttt{AS} clause, for referring to them mnemonically with respect to the application domain.
\end{itemize}

\noindent{\bf Discussion.}
This definition of semantics is quite close to the relational one, but some differences require further discussion.
\begin{itemize}
\item {\bf Choice of {\tt LABELS}.}
Adopting labels for identifying the trigger's target appears to be the most natural choice in the case of Property Graphs -- every node and relationship, sharply, either belongs or does not belong to the set of items with a given label. Still, the situation is more complex than in the relational case, as a node or relationship may have no label, or instead it may have more than one associated label, whereas a tuple belongs to exactly one table in the relational model.\footnote{Labels could be substituted by PG-Types (with {\tt STRICT} option, which makes them compulsory for all nodes or relationships) if they will become widely used and accepted as standard.}
For clarity of the execution semantics, we also make the assumption that {\it no trigger can monitor the setting or removal of its target label} and that
\textit{the target label cannot be set or removed within the \texttt{<statement>}}.
\item The {\tt ONCOMMIT} option, which is not supported in relational databases, is supported by Neo4j and Memgraph, although with several options and different semantics for each option (see Section \ref{sec:mapping}). Our interpretation of {\tt ONCOMMIT} is to consider and execute the trigger when the transaction reaches the commit point, and then include also triggers' side effects before actually committing. The {\tt ONCOMMIT} option is in part justified in graph databases by the interleaving of {\tt MATCH} clauses with node and relationship creations, updates and deletions, which make the context-of-operation less well-defined than in a relational database.\footnote{The {\tt ONCOMMIT} option is sometimes advocated for relational databases, e.g., in practitioners' blogs.} 
\item Similarly, for what concerns the order of execution of different triggers that are activated by the same Cypher or GQL query, note that these queries are generally much more powerful than relational update queries, targeted to a single table, as they can update multiple nodes and relationships, with different labels. Hence, the most sensible option for prioritizing them is to resort to the trigger creation time, providing a totally ordered prioritization.\footnote{Another option would be to use a total order based on the names of the triggers, as some relational databases do (e.g., PostgreSQL; see~\url{https://www.postgresql.org/docs/current/sql-createtrigger.html}).} In all cases, with a given execution order, the execution semantics of cascading triggers should mimic the relational one, with the stack of trigger execution contexts as described, e.g., in~\cite{melton2001sql}.
\end{itemize}


\section{Mapping PG-Triggers to Neo4j and Memgraph}
\label{sec:mapping}

In this section, we describe possible mapping strategies between PG-Triggers and Neo4j/Memgraph implementations.
Comparable mappings could be performed with the other NoSQL Graph Technologies with Event Listeners 
and programmatic support similar to APOC.  

\subsection{Mapping PG-Triggers to APOC triggers}
Triggers are not natively supported by Neo4j~\cite{robinson2015graph} and Cypher at the current time; however, we can make use of their implementation in the {\tt APOC} library,\footnote{We refer to Version 4.4, available as of May 2023; APOC triggers implementation is unchanged in Version 5.14, available as of November 2023}. created as a Neo4j community effort. The library includes over 450 procedures, providing functionalities for utilities, conversions, graph updates, data import, data transformations, and manipulations. 

We focus on the {\tt apoc.trigger} collection of procedures, which includes several trigger operations. Triggers can be created (\texttt{install}), deleted (\texttt{drop} or \texttt{dropAll}), paused (\texttt{stop}), and resumed (\texttt{start}); the creation of a trigger uses the following syntax:
\begin{center}   
\texttt{apoc.trigger.install(databaseName, name, statement, selector)}
\end{center}
The parameters refer to the database in use, the name of the trigger, its code (statement), and its action time (selector);\footnote{A fifth parameter {\tt config} refers to Neo4j configuration, left empty as it is not relevant for the trigger translation.} note that the specific trigger event (e.g., {\tt create, delete, set, remove}) is defined within the trigger code. The `selector' parameter indicates the time at which the trigger is activated. Four cases are supported:
\begin{itemize}
    \item \texttt{before}: the trigger activates right before the commit of the current transaction - this is our {\tt ONCOMMIT} option, and the default behavior;
    \item \texttt{rollback}: the trigger activates after the transaction is rolled back, within a new transaction;
    \item \texttt{after}: the trigger activates in a new transaction after the commit of the current one, but within the same thread; 
    \item \texttt{afterAsync}: the trigger activates in a new transaction after the commit of the current one; however, unlike the \texttt{after} phase, the execution occurs in a separate thread avoiding possibly blocking operations. 
\end{itemize}
The {\tt before} and {\tt after} action times of APOC triggers are discouraged by the APOC community, as they may create blocking 
conflicts with the query that causes the trigger activation; as a consequence, the advised action time is \texttt{afterAsync}. We have also experienced several unpredictable blocking conditions while testing the {\tt after} action time, and thus adopted the \texttt{afterAsync} option. Note, however, that such a pragmatic approach does not guarantee that triggers will see the final state produced by the transaction that activates them, since other transactions can occur after the commit of the activating transaction and before the trigger actually starts its execution. 

\begin{table}[b]
\resizebox{\linewidth}{!}{%
\begin{tabular}{ll}
\toprule
    \textbf{Statement} & \textbf{ Description}\\ 
\toprule
    \texttt{createdNodes} & list of created nodes\\ 
    \texttt{createdRels} &  list of created relationships\\ 
    \texttt{deletedNodes} & list of deleted nodes\\ 
    \texttt{deletedRels} & list of deleted relationships\\ 
    {\tt assignedLabels} & set of new labels for an item\\ 
    \texttt{removedLabels} & set of removed labels from an item\\ 
    {\tt assignedNodeProperties} & quadruple representing <target node,\\
    & property name, old value, new value>\\ 
    {\tt assignedRelProperties} & quadruple representing <target rel,\\
    & property name, old value, new value>\\ 
    \texttt{removedNodeProperties} & triple representing <target node,\\
    & property name, old value>\\ 
    \texttt{removedRelProperties} & triple representing <target rel,\\
    & property name, old value>\\
    \bottomrule
\end{tabular}
}
\caption{Utility functions in the APOC trigger procedures used to indicate the action type and capture the old and new states; note that {\tt Relationship} is shortened to {\tt Rel}}
\label{tab:parameters}
\end{table} 

\begin{figure*}[h!]
    \centering
\includegraphics[width=0.78\linewidth]{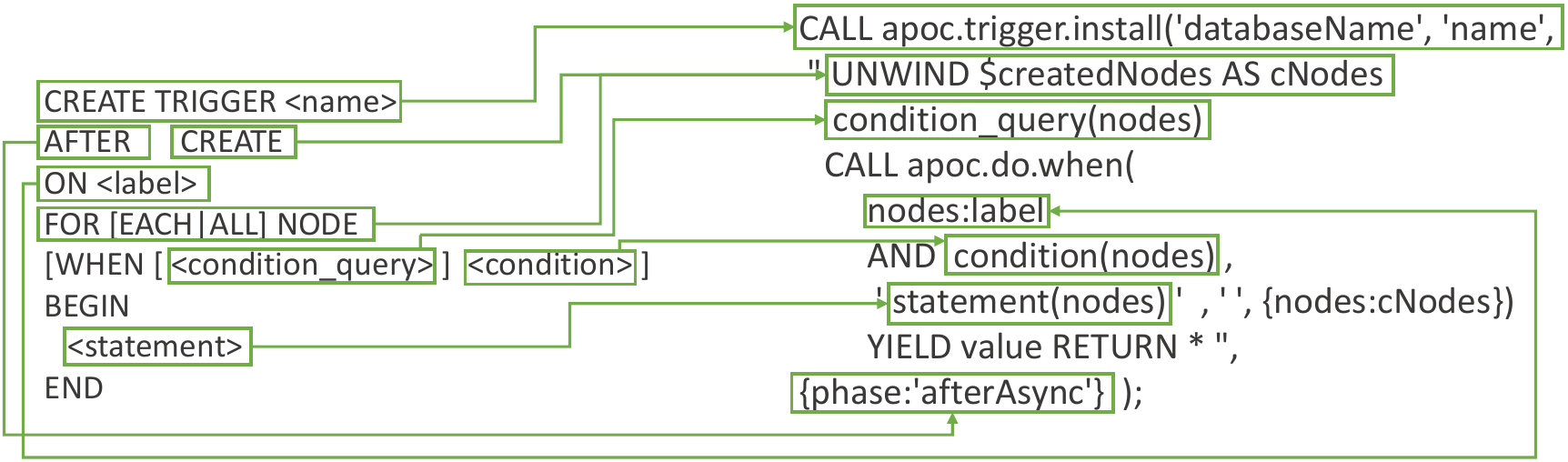}
    \caption{Syntax-directed translation from PG-Triggers to APOC triggers, for the specific case of node creation}
    \label{fig:apocMapping}
\end{figure*}

\begin{table*}
\resizebox{0.8\linewidth}{!}{%
\begin{tabular}{cccc}
\toprule
 &  & \textbf{OLD} & \textbf{NEW} \\ \toprule
\multirow{2}{*}{Nodes} & {Create} & {-} & \$createdNodes \\ \cmidrule(r){2-2} \cmidrule(l){3-4} 
 & Delete & \$deletedNodes & {-} \\ \midrule
\multirow{2}{*}{Relationships} & {Create} & {-} & \$createdRelationships \\  \cmidrule(r){2-2} \cmidrule(l){3-4} 
 & Delete & {\$deletedRelationships} & {-} \\ \midrule
\multirow{2}{*}{Labels} & Set & {-} & {\$assignedLabels} \\  \cmidrule(r){2-2} \cmidrule(l){3-4} 
 & {Remove} & {\$removedLabels} & {-} \\ \midrule
Node / Rel & Set & \$assignedProperties(node, property, old, -) & {\$assignedProperties(node, property, -, new)} \\  \cmidrule(r){2-2} \cmidrule(l){3-4} 
Properties  & {Remove} & {\$removedProperties(node, property, old)} & {-}\\
\bottomrule
\end{tabular}
}
\caption{Syntax-directed scheme for building {\tt OLD} and {\tt NEW} transition variables in Neo4j}
\label{tab:oldnew}
\end{table*}

In the current implementation of the APOC library, APOC triggers do not cascade (e.g., recursively activate) correctly. Specifically, in the \texttt{before} case, all the installed triggers are activated, only once, in alphabetic order, regardless of the specific node or relationship type that is monitored; thus, the sequence of activation dictates whether one trigger can react, just for one time, to the events produced by the other one. On the other end, in the \texttt{after} and {\tt afterAsynch} cases all triggers are executed within a single, new transaction launched after the original one, but a cascade of triggers is explicitly blocked by an initial clause that excludes considering data generated by other triggers (this information is carried within associated metadata). This severe limitation must be overcome in future releases so as to guarantee relevant applications of triggers over graphs, such as inferring properties of paths of arbitrary length.

Figure~\ref{fig:apocMapping} shows the syntactic mapping between a PG-Trigger reacting to {\it node creation} and the corresponding {\it APOC trigger install};
note that similar syntax-directed translations can be easily drafted for all ten kinds of supported events 
 \{{\it node, relationship}\} $\times$ \{{\it creation, deletion}\} and 
\{{\it label, node-property, relationship-property}\} $\times$ \{{\it set, removal}\}.

As discussed, the APOC {\tt install/drop} procedures have four parameters: the {\tt databaseName} (not present on the left side), the 
trigger \texttt{name}
(copied from the left side), the 
{\tt statement} and the 
{\tt} selector (i.e., action time `\texttt{afterAsync}').
The richest parameter is the {\tt statement}, as it, in turn, includes a pair of standard statements. 

\begin{itemize}
\item The first one is a call to the {\tt UNWIND} clause,
returning the list of items affected by the event (in the example, the list of created nodes); these are renamed (as {\tt cNodes}) so as to become usable in the rest of the statement. Table~\ref{tab:parameters} shows the different trigger procedures that can be intercepted by {\tt UNWIND} to capture the different action types.
\item Then, in our translation scheme, we opted for using the APOC {\tt do.when} procedure, so as to generate placeholders for the condition as well as the action part of the trigger. 
\end{itemize}
The {\tt do.when} procedure has four parameters: 
the condition, 
the action if the condition \textit{is} met, 
the action if the condition \textit{is not} met, and
the operands that can be used in the condition and action.
The procedure returns (\texttt{YIELD}s) a value.
In our translation, the {\tt do.when} condition is a conjunctive predicate extracting the items with a given label or property (taken from the left side) which satisfies the \texttt{condition} predicate (also taken from the left side). In APOC, the condition is a Boolean expression of simple terms; if terms need to be extracted from the data graph, it is necessary to place a \texttt{condition\_query} before the \texttt{do.when} procedure.

The first {\tt do.when} action is executed when the condition is true; it uses the trigger \texttt{statement}
code (taken from the left side); the second action, executed when the condition is false, is an empty string since nothing needs to be done in that case. Both the \texttt{condition} predicate and the trigger \texttt{statement}
refer to specific nodes, and these are the \texttt{cNodes} created by the {\tt UNWIND} clause and appearing as fourth {\tt do.when} parameter.   

A number of additional aspects can be noted, as explained in the following. First, we cannot separate the two cases of granularity (item or set oriented), because the {\tt UNWIND} clause returns, in any case, the entire set of involved items.\footnote{{\tt UNWIND} returns a list rather than a set, but there is no definition of the order in which the list is produced.} Thus, the \texttt{<statement>} is in charge of considering, within its code, either each item or, collectively, the set of all items.
Second, the utility functions in Table~\ref{tab:oldnew} also allow us to build the {\tt OLD} and {\tt NEW} transition variables for all supported events; these refer either to each
individual transition value or to the set of transition values, as the supported procedures do not distinguish the two cases. Note that each transition variable is appropriately paired to events, e.g., {\tt NEW} transition variables are defined for the creation of nodes and relationships and for the setting of new labels and properties.  

Several examples of the translations of various trigger options, including the use of {\tt OLD} and {\tt NEW} variables, are discussed in the examples in Section~\ref{sec:example}. 

\begin{center}
    \begin{figure*}[h]
    \centering
    \includegraphics[width=0.78\linewidth]{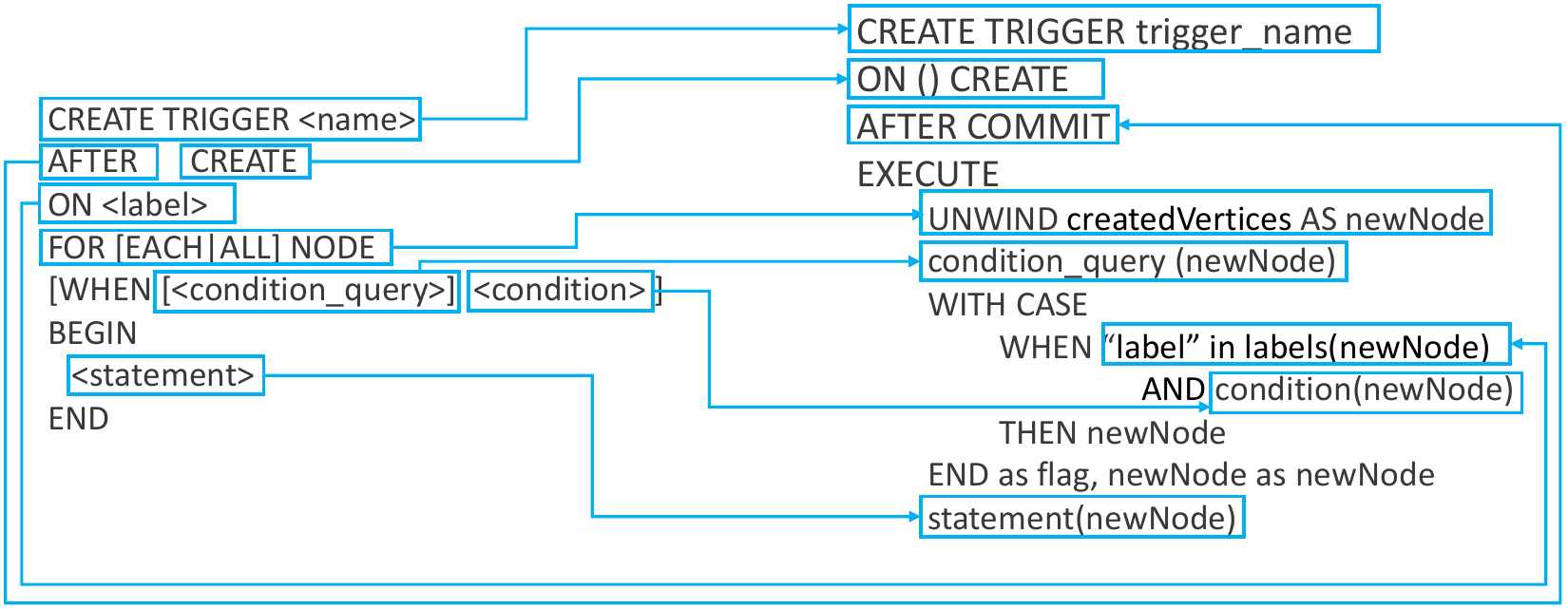}
    \caption{Syntax-directed translation from PG-Triggers to Memgraph triggers, for the specific case of node creation.} 
    \label{fig:memgraphMapping}
\end{figure*}
\end{center}

\subsection{Mapping PG-Triggers to Memgraph triggers}

Memgraph offers a direct implementation of triggers that embeds any legal openCypher query.
To create a new trigger the following syntax is used:
\begin{Verbatim}
CREATE TRIGGER trigger_name 
[ ON [ () | --> ] CREATE | UPDATE | DELETE ]
[ BEFORE | AFTER ] COMMIT
EXECUTE openCypherStatements
\end{Verbatim}

\begin{table}
\resizebox{\linewidth}{!}{%
\begin{tabular}{ll}
\toprule
    \textbf{Variable} & \textbf{ Description}\\ 
\toprule
    \texttt{createdVertices} & list of created nodes\\
    \texttt{createdEdges} & list of created relationships\\
    \texttt{createdObjects} & list of created objects (as maps)\\
    \texttt{updatedVertices} & list of node updates \\
    &(set/removed properties/labels)\\
    \texttt{updatedEdges} & list of node updates \\
    &(set/removed properties) \\
    \texttt{updatedObjects} & list of node/rels updates \\&(set/removed properties/labels)\\
    \texttt{deletedVertices} & list of deleted nodes\\
    \texttt{deletedEdges} & list of deleted relationships\\
    \texttt{deletedObjects} & list of deleted objects (as maps)\\
     \texttt{setVertexLabels} & list of set node labels\\    
      \texttt{removedVertexLabels} & list of removed node labels \\
    \texttt{setVertexProperties} & list of set node properties\\
    \texttt{setEdgeProperties} & list of set relationship properties\\
   \texttt{removedVertexProperties} & list of removed node properties\\
    \texttt{removedEdgeProperties} & list of removed relationship prop.\\   
    \bottomrule
\end{tabular}
}
\caption{Predefined variables for Memgraph triggers capturing the old and new states.}
\label{tab:memgraph_parameters}
\end{table} 

This syntax allows specifying the trigger name, the event type, and the execution time; then,
the trigger syntax allows the inclusion of any valid openCypher query/statement, to be executed on activation.\footnote{OpenCypher is very similar to Cypher, with differences discussed in \cite{diffcypheropen}.} Event types discern among the creation, update, or deletion of nodes (denoted by the symbol {\tt ()} and named {\it vertices}), relationships (denoted by the symbol {\tt --->} and named edges), or any such object (objects include both vertices and edges).
Action times include {\tt before commit} and {\tt after commit}; in the former case the statement is executed right before the commit of the current transaction - along our {\tt ONCOMMIT} option; in the latter case the statement is executed asynchronously after the transaction is committed.  As in Neo4j APOC triggers, several predicates allow capturing the new vs old states relative to the given event type, see Table \ref{tab:memgraph_parameters}. 
The trigger management implementations of {\tt before commit} and {\tt after commit} in Memgraph are identical to those of Neo4J APOC procedures, therefore also in Memgraph triggers do not correctly cascade. 

Figure~\ref{fig:memgraphMapping} shows the syntactic mapping between a PG-Trigger reacting to {\it node creation} and the corresponding Memgraph trigger creation;
note that similar syntax-directed translations can be easily drafted for all fifteen kinds of supported events \{{\it vertex, edge, object}\} $\times$ \{{\it creation, update, deletion}\} and \{{\it label, vertex-property, edge-property}\} $\times$ \{{\it set, removal}\}.
Notably, Figures~\ref{fig:apocMapping} and~\ref{fig:memgraphMapping} offer similar translation strategies, except for a few syntactical differences. For instance, note that Memgraph moves all the logic inside the openCypher statement; accordingly, while in the Neo4j APOC implementation we used the APOC \texttt{do.when} function to express conditional execution, in Memgraph we can express the condition with openCypher's CASE construct. 


%
    
Note also that variables computed by the \texttt{condition\_query} must be propagated to the \texttt{statement} part of the query by specifying aliases for them in the \texttt{WITH} clause. 
The final statement can be of arbitrary complexity in openCypher; in order for the condition expressed in the \texttt{CASE} to be effective, the execution of the statement must include as early condition the predicate "\texttt{WHERE flag is not NULL}" and its execution can progress only if the predicate is true.

\begin{figure*}[h]
    \centering
    \includegraphics[width=\linewidth]{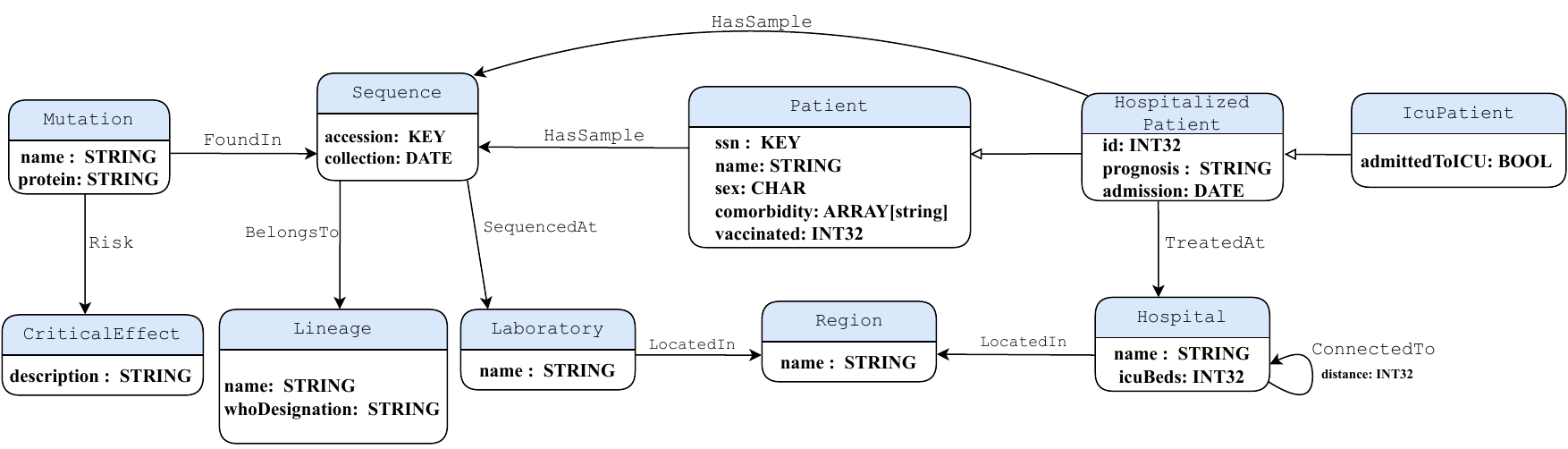}
    \caption{PG-Schema for the graph database used in our running example}
    \label{fig:schema}
\end{figure*}

\begin{figure}[h]
\centering
\frame{\includegraphics[width=\linewidth]{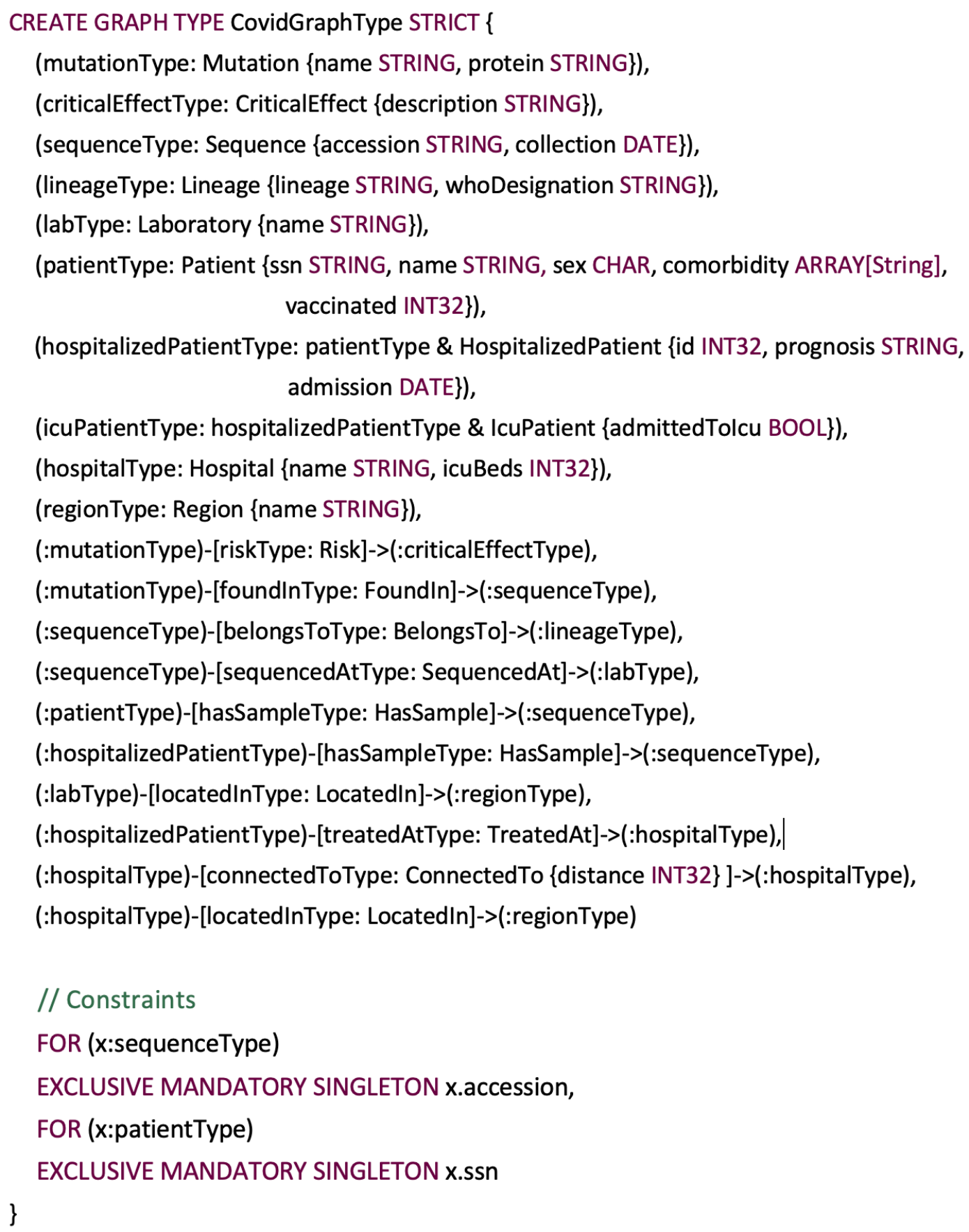}}
\caption{PG-Schema specification for the running example}
\label{fig:pgschema}
\end{figure}

\section{Running example: COVID-19 and SARS-CoV-2}
\label{sec:example}
Since the outbreak of the COVID-19 pandemic, the disease evolution has been constantly monitored; meanwhile, SARS-CoV-2, the virus responsible for the disease, has continually mutated its genomic sequence; thanks to the evolutive selection, the virus has developed increased transmissibility, host immune evasion, and resistance to antivirals~\cite{volz2021assessing,focosi2021analysis}. The availability of genome sequences collected over time has been very useful for molecular surveillance of the epidemic and for the evaluation and planning of effective control strategies. This scenario is sufficiently rich and diversified to illustrate the expressivity and versatility of our PG-Triggers proposal.

\subsection{PG-Schema}

We designed several data models for dealing with COVID-19 and SARS-CoV-2~\cite{bernasconi2020empowering,bernasconi2021review}; in particular, we proposed the {\it CoV2K knowledge base}~\cite{alfonsi2022cov2k}, providing a description of SARS-CoV2 sequences, of their amino acid mutations and their effects, of the clustering of sequences within lineages and variants, and of the assignment of sequences to donors (i.e., patients). We next model an excerpt of CoV2K as a graph database; in particular, we use the abstractions from the PG-Schema proposal~\cite{bonifati2022pg} so as to take advantage of its rich semantics for graph definition. Note that, in PG-Schema, all nodes are typed and have a unique label;
the schema also supports type hierarchies (e.g., between {\tt  Patient} and {\tt HospitalizedPatient}), with inheritance. Constraints, including keys, are separately defined. 

The adoption of PG-Schema makes graph databases more similar to relational databases, especially with a {\tt STRICT} graph type definition, where labels uniquely identify nodes in the same way as table names identify relational tables; in this context, changing labels is not possible, thereby implicitly satisfying the semantic constraint on legal statements introduced with the PG-Trigger semantics (see Section~\ref{sub:trigger_semantics}), 
which disallowed setting or removing, in the \texttt{statement}, labels defining the target.
Note also that relationships are implicitly identified not only by their types but also by the types of nodes that they connect. However, our standard proposal supports generic graph databases, where nodes and relationships can have multiple labels and some of them can be unlabeled.

The PG-Schema of our running example is shown in Figure~\ref{fig:schema}. It includes: \texttt{Mutation} with \texttt{name} (e.g., {\tt "Spike:D614G"}) and \texttt{protein} (e.g., {\tt "Spike"}), their relationship  with \texttt{CriticalEffect}s (with their \texttt{description}, e.g., {\tt "Enhanced infectivity"}). \texttt{Sequence}s are characterized by their \texttt{accession} (key) and the relationship with their mutations; each \texttt{Sequence} belongs to a \texttt{Lineage}, for which we know the \texttt{name} and an optional \texttt{whoDesignation} (a property assigned by the World Health Organization, e.g., {\tt "Alpha"}). \texttt{Sequence}s are collected at a given date of \texttt{collection}, within \texttt{Laboratory}s, which belong to \texttt{Region}s; they are sampled from \texttt{Patient}s. 

\sloppy{\texttt{Patient}s have an \texttt{ssn} (a key), \texttt{name}, \texttt{sex}, and a set of \texttt{comorbidity} values (e.g., {\tt "diabetes"}). Some patients are \texttt{vaccinated}; in this case, the type is INT32 to denote the number of vaccinations (0 if the patient is not vaccinated, else the number of vaccine shots). Some \texttt{Patient}s can be admitted to \texttt{Hospital}s, which are \texttt{name}d and located within \texttt{Region}s; each \texttt{Hospital} has a maximum number of intensive care beds (\texttt{icuBeds}), and pairs of \texttt{Hospital}s are connected by given distances. On admission, \texttt{HospitalizedPatient}s are associated with an internal \texttt{id} and a \texttt{prognosis} (e.g., {\tt "severe"}); some of them (the \texttt{IcuPatient}s) may also be admitted to intensive care units, on given admission dates. 
The PG-Schema specification for this graph database is shown in Figure~\ref{fig:pgschema}.}

\subsection{PG-Triggers}

We next present several PG-Triggers that progressively illustrate the various features of the proposed standard. In order to write efficient queries, the code should normally refer to transition variables, which are the {\it handlers} to the part of the graph that has been modified. The first five triggers produce {\it alert nodes}, i.e., nodes that -- once described with PG-Schema -- are of a new, {\tt OPEN} type (allowing for the inclusion of arbitrary properties).
All the alerts include the 
\texttt{time}
when they are produced and a textual
{\tt desc}ription.

\subsubsection{Simple reactions to node, relationship, and property creation}\label{sec:simple}

The first trigger reacts to the fact that a new mutation is associated with a critical effect by creating an alert with the name of the mutation.

\begin{Verbatim}[frame=single]
CREATE TRIGGER NewCriticalMutation
AFTER CREATE
ON 'Mutation'
FOR EACH NODE
WHEN EXISTS (NEW)-[:Risk]-(:CriticalEffect)
BEGIN
    CREATE (:Alert{time:DATETIME(), 
        desc:'New critical mutation', 
        mutation:NEW.name}) 
END
\end{Verbatim}

The second trigger, similar to the previous one, reacts to the association of a critical mutation with a lineage (i.e., a viral subspecies, also informally called {\it variant}) and creates an alert for the lineage. Note that in this example the condition part is merged within the action; as in relational triggers, the separation between condition and action may be arbitrary. 

\begin{Verbatim}[frame=single]
CREATE TRIGGER NewCriticalLineage
AFTER CREATE
ON 'BelongsTo'
FOR EACH RELATIONSHIP
WHEN
  MATCH (s:Sequence)-[NEW]-(l:Lineage)
  WHERE EXISTS { 
    MATCH (:CriticalEffect)-[:Risk]- 
    (:Mutation)-[:FoundIn]-(s) 
    }
BEGIN
    CREATE (:Alert{time:DATETIME(), 
        desc:'New critical lineage', 
        lineage:l.name}) 
END
\end{Verbatim}

The third trigger monitors a simple change in the \texttt{whoDesignation} property, e.g., the change of {\tt Indian} to {\tt Delta}:

\begin{Verbatim}[frame=single]
CREATE TRIGGER WhoDesignationChange
AFTER SET 
ON 'Lineage'.'whoDesignation'
FOR EACH NODE
WHEN OLD.whoDesignation <> NEW.whoDesignation
BEGIN
    CREATE (:Alert{time: DATETIME(), 
    desc:'New Designation for an existing Lineage'})
END
\end{Verbatim}

\subsubsection{Conditions using fixed thresholds vs state comparisons}
\label{sec:max-thresh-incr}

The next trigger counts the patients who require intensive care at the Sacco Hospital and raises an alert when their number exceeds 50 patients. Note the use of two labels to denote 
matching along type hierarchies.
Note also that the trigger uses set granularity (\texttt{FOR ALL}) and no transition variable is needed, as the counting relates to all the patients.

\begin{Verbatim}[frame=single]
CREATE TRIGGER IcuPatientsOverThreshold
AFTER CREATE
ON 'IcuPatient'
FOR ALL NODES
WHEN
  MATCH (p:HospitalizedPatient:IcuPatient)
    -[:TreatedAt]-(:Hospital{name:'Sacco'})
  WITH COUNT(p) AS icuPat 
  WHERE icuPat > 50
BEGIN
   CREATE (:Alert{time:DATETIME(),desc:'ICU patients 
   at Sacco Hospital are more than 50'})
END
\end{Verbatim}

The next trigger compares the patients who are in intensive care at the Sacco Hospital after admission, and raises an alert when the new patients in ICU are more than 10\% of the total of patients in ICU; we assume that admissions are periodically registered by a transaction, e.g., daily.  Note that this trigger also uses set granularity and the transition variable \texttt{NEWNODES} is used in the comparison; as we assume that transition variables correspond to a well-defined set of nodes affected by the event, we can further define new variables over them.

\begin{Verbatim}[frame=single,fontsize=\small]
CREATE TRIGGER IcuPatientIncrease
AFTER CREATE
ON 'IcuPatient'
FOR ALL NODES
WHEN
 MATCH (p:HospitalizedPatient:IcuPatient)-
    [:TreatedAt]-(:Hospital{name: 'Sacco'}),
 MATCH (pn:NEWNODES)-[:TreatedAt]-(:Hospital{name:'Sacco'})
 WITH  COUNT(pn) AS NewIcuPat,
        COUNT(p) AS TotalIcuPat
 WHERE NewIcuPat / TotalIcuPat > 0.1
BEGIN 
 CREATE (:Alert{time:DATETIME(),desc:'ICU patients 
 at Sacco Hospital have increased by > 10%'})
END
\end{Verbatim}

\subsubsection{Triggers with side effects in the action}\label{sec:relocation}

The next trigger describes the relocation of patients from the Sacco Hospital (in Lombardy) to the Meyer Hospital (in Tuscany), caused by the unavailability of ICU beds.\footnote{Patients' relocations out of Lombardy occurred during the first pandemic wave, not only to Tuscany but also to Germany and Switzerland.} 
Note that the trigger evaluates, in the condition, if ICU beds at Sacco are insufficient due to the new admissions; if so, the statement first considers the availability of ICU beds at the Meyer Hospital. If they are sufficient for hosting the new admissions at Sacco, then these patients are moved from the Sacco to the Meyer Hospital. Note that the patient relocation is rendered in the graph database by removing, for each patient, the relationship with the Sacco Hospital, and adding the relationship with the Meyer Hospital. 

\begin{Verbatim}[frame=single, fontsize=\small]
CREATE TRIGGER IcuPatientMove
AFTER CREATE
ON 'IcuPatient'
FOR ALL NODES
WHEN 
  MATCH (p:HospitalizedPatient:IcuPatient)-[:TreatedAt]- 
    (h:Hospital{name:'Sacco'}) 
  WITH COUNT(p) AS TotalIcuPat 
  WHERE TotalIcuPat > h.icuBeds 
BEGIN 
  MATCH(pn:NEWNODES)-[:TreatedAt]-(:Hospital{name:'Sacco'}),
  MATCH(pt:HospitalizedPatient:IcuPatient)-[:TreatedAt]-
    (ht:Hospital {name:'Meyer'})
  WITH COUNT(pt) AS MeyerICU, ht.IcuBeds AS MeyerBeds,
       COUNT(pn) AS newICUSacco
  WHERE newICUSacco + MeyerICU <= MeyerBeds 
  THEN FOREACH (p IN pn) 
  BEGIN
    MATCH (p)-[c:TreatedAt]-(:Hospital{name:'Sacco'})
    DELETE c
    CREATE (p)-[:TreatedAt]->(:Hospital{name:'Meyer'})
  END
END
\end{Verbatim}

The final example gives a different solution to the same problem, i.e., reacting to the unavailability of ICU beds in any Hospital in the Lombardy Region.
Note that this trigger is at the item level, operates upon all hospitals in Lombardy where there are new patients admitted to ICU, and moves newly admitted patients from those hospitals where ICU beds are exceeded, by finding the closest hospital where patients may be relocated. Note that patients of the same hospital are moved together to the closest hospital (not necessarily in `Lombardy'). 

\begin{Verbatim}[frame=single,fontsize=\small]
CREATE TRIGGER MoveToNearHospital
AFTER CREATE
ON 'IcuPatient'
FOR EACH NODE
WHEN 
  MATCH (NEW:HospitalizedPatient:IcuPatient)
    -[:TreatedAt]-(h:Hospital)
    -[:LocatedIn]-(:Region{name:'Lombardy'}),
  MATCH (p:IcuPatient)-[:TreatedAt]-(h)
  WITH COUNT(p) AS TotalIcuPat, h
  WHERE TotalIcuPat > h.icuBeds
BEGIN 
  MATCH (h:Hospital)
    -[:LocatedIn]-(:Region{name:'Lombardy'}),
  MATCH (pn:NEW)-[:TreatedAt]-(h)
    -[ct:ConnectedTo]-(hc:Hospital)
  WITH ct ORDER BY ct.distance LIMIT 1
  THEN 
    BEGIN
     MATCH (pn)-[c:TreatedAt]-(h)
     DELETE c
     CREATE (pn)-[:TreatedAt]->(hc)
    END
  END
\end{Verbatim} 

Note that this trigger may not converge if ICU beds in close hospitals are also exceeded, as the first trigger execution could cause an infinite cascade of trigger executions. Termination analysis for triggers is a well-known research topic, and in particular, by using the methods discussed in~\cite{baralis1994better} one could prove that recursion terminates when the availability of beds is tested prior to moving patients, while failure to do the test may lead to potential non-termination.

\subsection{Translation to APOC triggers}
We consider four examples and present their translation to Neo4j APOC triggers;
a translation to Memgraph would follow the same principles and therefore is omitted.
In general, the translation is quite readable, although there are many void parameters due to the specific syntax of the APOC trigger procedures.

Note that type hierarchies are not supported in Neo4j, thus we model the hierarchy from {\tt HospitalizedPatient} to {\tt IcuPatient} as a conventional {\tt Isa} relationship, directed from the subtype to the type.
Note also that the code of APOC triggers should be coherent with the trigger granularity: with {\it item granularity} the condition predicate and triggered statement should be item-based; with {\it set granularity} they should be set-based and in particular include aggregate predicates. Also, recall that the utilities of Table~\ref{tab:parameters} return an arbitrarily ordered list of all the transition variables for the given transaction, regardless of its granularity. 

The trigger {\tt WhoDesignationChange} of Subsection~\ref{sec:simple} includes a condition using both the {\tt OLD} and {\tt NEW} transition variables. Note that the {\tt UNDWIND} clause is exploited in order to extract all the terms used in the Boolean condition, taking advantage of the hierarchical structure of the {\tt \$assignedNodeProperties} parameter present in the first argument of the {\tt do.when} APOC procedure, thus no condition query is necessary. Its translation is:

\begin{Verbatim}[frame=single,fontsize=\small]
CALL apoc.trigger.install('databaseName', 
                          'WhoDesignationChange', 
"UNWIND keys($assignedNodeProperties) AS k
 UNWIND $assignedNodeProperties[k] AS aProp
 WITH aProp.node AS node, collect(aProp.key) AS propList, 
    aProp.old as oldValue, aProp.new as newValue
 CALL apoc.do.when(
    node:Lineage AND 'whoDesignation' IN propList
                 AND oldValue <> newValue, 
    'CREATE (:Alert{time: DATETIME(), 
     desc: \"New Designation for an existing Lineage\"})',
     '',{})
 YIELD value RETURN *",
{phase:'afterAsync'});
\end{Verbatim}


The trigger {\tt IcuPatientIncrease} of Section~\ref{sec:max-thresh-incr} demonstrates the need to use {\tt Isa} relationships as a replacement for type hierarchies. Note, in this case, the use of the condition query in order to extract the terms used in the Boolean condition of the first argument of the {\tt do.when} APOC procedure. 

\begin{Verbatim}[frame=single,fontsize=\small]
CALL apoc.trigger.install('databaseName', 
                          'IcuPatientIncrease', 
"UNWIND $createdNodes as cNodes
 MATCH (p:IcuPatient)-[:Isa]-(:HospitalizedPatient)
    -[:TreatedAt]-(h:Hospital{name:'Sacco'})
    WITH COUNT(cNodes) AS NewIcuPat,
         COUNT (p) AS TotalIcuPat, cNodes
 CALL apoc.do.when(
    cNodes:IcuPatient AND  NewIcuPat/TotalIcuPat > 0.1,
    'MERGE (:Alert{time:DATETIME(), desc:\"ICU patients 
    at Sacco Hospital have increased more than 10%\"})', 
    '', {} )
 YIELD value RETURN *",
{phase:'afterAsync'});
\end{Verbatim}

The trigger {\tt IcuPatientMove} of Section~\ref{sec:relocation}
uses the condition so as to check whether patients at Sacco exceed the number of ICU beds; the statement implements the patient transfer, subject to availability at Meyer Hospital. Note that the pair of creation and deletion of relationships is done using two separate {\tt FOREACH} 
conditions; note also that a condition query is required in this case.

\begin{Verbatim}[frame=single,fontsize=\small]
CALL apoc.trigger.install('databaseName',
                          'IcuPatientMove',
"UNWIND $createdNodes AS cNodes
 MATCH (:IcuPatient)-[:Isa]-(p:HospitalizedPatient)-
    [:TreatedAt]-(h:Hospital{name:'Sacco'})
 WITH COUNT(p) AS TotalIcuPat, 
      h.IcuBeds AS TotalBeds, 
      cNodes
 CALL apoc.do.when(
   cNodes:IcuPatient AND TotalIcuPat > TotalBeds,
   'MATCH (pt:IcuPatient)-[:Isa]-(:HospitalizedPatient)
       -[:TreatedAt]-(ht:Hospital{name:$Meyer}) 
   WITH COUNT(pt) AS MeyerICU, ht.IcuBeds AS MeyerBeds,
          COUNT(cNodes) AS newICUSacco, ht, cNodes
   WHERE newICUSacco + MeyerICU <= MeyerBeds 
   MATCH (cNodes)-[:Isa]-(:HospitalizedPatient)
       -[c:TreatedAt]-(:Hospital{name:$Sacco})
   FOREACH (p IN [cNodes] | DELETE c)
   FOREACH (p IN [cNodes] | CREATE(p)-[:TreatedAt]->(ht))',   
   '', {cNodes:cNodes, Meyer:'Meyer', Sacco:'Sacco'})
 YIELD value RETURN count(*)",
{phase: 'afterAsync'});
\end{Verbatim}

Finally, we present the trigger {\tt MoveToNearHospital} of Section~\ref{sec:relocation}.

\begin{Verbatim}[frame=single, fontsize=\small]
CALL apoc.trigger.install('databaseName', 
                          'MoveToNearHospital', 
"UNWIND $createdNodes AS cNodes
 MATCH (cNodes)
  -[:Isa]-(:HospitalizedPatient) 
  -[:TreatedAt]-(h:Hospital)
  -[:LocatedIn]-(:Region{name:'Lombardy')
 MATCH (:IcuPatient)-[:Isa]-(p:HospitalizedPatient)
   -[:TreatedAt]-(h)
 WITH COUNT(p) AS TotalIcuPat, 
      h.icuBeds AS TotalIcuBeds,
      cNodes, h
 CALL apoc.do.when(
   nodes:IcuPatient AND TotalIcuPat > TotalIcuBeds,
   'MATCH (h)-[ct:ConnectedTo]-(hc:Hospital)
    WITH ct, cNodes, h, hc ORDER BY ct.distance ASC LIMIT 1
    MATCH (cNodes)-[:Isa]-(ph:HospitalizedPatient)
           -[c:TreatedAt]-(h)
    FOREACH (pat in [cNodes] | DELETE c)
    FOREACH (pat in [cNodes] | CREATE (ph)
            -[:TreatedAt]->(hc))',
   '',{CNodes:cNodes, h:h}) 
 YIELD value RETURN *",
{phase: 'afterAsync'});
\end{Verbatim}

\section{Discussion and Conclusion}
\label{sec:conclusion}

In this article, we have shown that
adding reactive components to graph databases is at the same time very natural, along the SQL3 standard, and also very useful, as reactive programming
can leverage graph databases in supporting important applications; one of them, highlighted in our running example, is the management of clinical emergencies due to new mutations and lineages (also known as variants) during the COVID-19 pandemic. 
We have shown that the concepts of PG-Triggers descend naturally from relational concepts, although they require adaptation due to the richer model of graph databases, which includes nodes, relationships, labels, and properties. 

We have also shown that PG-Triggers can be supported on top of Neo4j by making use of the APOC trigger procedures, with straightforward syntax-directed translations - once the various ingredients have been understood and mastered. A major drawback is that APOC triggers miss some important ingredients for being fully compliant with standardization needs, including the mastering of activation before or after operations that cause triggering and a complete and correct management of cascading changes. Moreover, being community-driven, APOC triggers are subject to changes - we experienced some changes during the development of our translation schemes. One objective of this article is also to motivate graph database companies, such as Neo4j, to support triggers as part of their standard offer.

\vspace{-2mm}
\section*{Resources}
A prototype of the Neo4j database, with scripts for data creation, population, and APOC triggers, 
is in the GitHub repository~\cite{repositorygithub,repositoryzenodo}.

\vspace{-2mm}
\begin{acks}
We thank Angela Bonifati for providing advice during the early phase of this research, and Ioana Manolescu for useful discussions on reactive knowledge management.
This paper is supported by FAIR (Future Artificial Intelligence Research) project, funded by the NextGenerationEU program within the PNRR-PE-AI scheme (M4C2, Investment 1.3, Line on Artificial Intelligence).
This work is part of Davide Magnanimi's Executive PhD Program at Politecnico di Milano.
\end{acks}

\balance
\printbibliography


\end{document}